\documentclass[aps,prd,superscriptaddress,twocolumn]{revtex4-1}
\usepackage{amssymb}
\usepackage{amsmath}
\usepackage{wallpaper}
\usepackage{booktabs}
\usepackage{comment}
\usepackage{bm}
\usepackage{multirow}
\usepackage{epsfig}
\def\be{\begin{equation}}
\def\ee{\end{equation}}
\def\bea{\begin{eqnarray}}
\def\eea{\end{eqnarray}}
\def\ba{\begin{equation}\begin{aligned}}
\usepackage{graphicx}

\usepackage[colorlinks,linkcolor=blue, urlcolor=blue, anchorcolor=blue, citecolor=blue]{hyperref}

\begin{document}

\title{Bayesian constraints on quark stars from multi-messenger observations}

\author{Wen-Jie Xie}
\email{xiewenjie@ycu.edu.cn}
\affiliation{Department of Physics, Yuncheng University, Yuncheng 044000, China}
\affiliation{Shanxi Province Intelligent Optoelectronic Sensing Application Technology Innovation Center, Yuncheng University, Yuncheng 044000, China}

\author{Cheng-Jun Xia}
\email{cjxia@yzu.edu.cn}
\affiliation{Center for Gravitation and Cosmology, College of Physical Science and Technology, Yangzhou University, Yangzhou 225009, China}
\author{Chen Zhang}
\email{zhangvchen@tongji.edu.cn}
\affiliation{School of Physics Science and Engineering, Tongji University, Shanghai 200092, China}
\affiliation{The HKUST Jockey Club Institute for Advanced Study, The Hong Kong University of Science and Technology, Hong Kong, P. R. China}

\author{Renxin Xu}
\email{r.x.xu@pku.edu.cn}
\affiliation{School of Physics and KIAA, Peking University, Beijing 100871, P. R. China}

\date{\today}
\begin{abstract}
We perform a systematic Bayesian analysis of quark star equations of state under current multimessenger constraints, investigating the impact of prior assumptions and extreme-mass observations. Quark matter is modeled within an interacting MIT bag framework that consistently accommodates color-superconducting phases (2SC, 2SC+s, and CFL) and perturbative QCD corrections. We find that quark star models exhibit a distinct advantage in naturally accommodating the ultra-low mass object HESS J1731-347, a configuration that is challenging for standard neutron star models. In the high-mass regime, the interpretation of the secondary component of GW190814 is shown to be strongly prior-dependent: only broad priors allow for the substantial stiffness required to support such a massive object ($\sim$2.6 M$_\odot$), while more restrictive priors favor a softer equation of state consistent with standard pulsar populations. Microscopically, we demonstrate that current data tightly constrain the effective bag constant and the overall stiffness, but cannot distinguish between different color-superconducting phases. Furthermore, we validate a reduction of the model to two effective parameters without loss of information. Our results indicate that if quark stars exist, their sound speeds consistently exceeds the conformal limit ($c_s^2>1/3$) at stellar densities.
\end{abstract}

\maketitle

\section{\label{sec:intro}Introduction}
Compact stars provide a unique laboratory for exploring the properties of matter under extreme densities, where the interplay between strong interactions and gravity becomes decisive. With typical masses of order (1-2 M$_\odot$) and radii of about 10 km, their central densities can exceed several times the nuclear saturation density ($n_0 \simeq 0.16~\mathrm{fm}^{-3}$), well beyond the regime accessible in terrestrial experiments. At such densities, the composition of matter inside compact stars remains highly uncertain, and the possible appearance of deconfined quark matter constitutes one of the central open questions in nuclear astrophysics \cite{Lattimer2004_Science304-536,Haensel2007_springer-book,Most2019_Phys.Rev.Lett.122-061101,Bauswein2019_PRL122-061102,Annala2020,Weih2020PRL124-171103,Yuan2025_PRD112-023019}.

Traditionally, compact stars have been modeled as neutron stars composed of charge-neutral and $\beta$-equilibrated hadronic matter. However, at sufficiently high densities, a transition to quark matter may occur, giving rise to hybrid stars~\cite{Annala2023_NC14-8451,Tang2025_PRD112-083009,Yuan2025_PRD112-023019} or, in more extreme scenarios, self-bound quark stars. In particular, the hypothesis that strange quark matter may represent the true ground state of quantum chromodynamics has motivated extensive studies of strange quark stars and related configurations \cite{Witten1984,FarhiJaffe1984,Alcock1986,Bodmer1971_PRD4-1601,Terazawa1979,Collins1975_PRL34-1353}. Depending on the underlying microphysics, quark matter may exist in different color-superconducting phases, such as the two-flavor superconducting (2SC) phase and the color–flavor-locked (CFL) phase, which can modify the equation of state (EOS) at high densities \cite{Alford2008}.

A key observational distinction between gravity-bound neutron stars and self-bound quark stars lies in their global structure. Neutron stars possess a vanishing density at the surface and therefore exhibit a minimum mass, whereas quark stars remain bound by strong interactions and can, in principle, exist with arbitrarily small masses \cite{Alcock1986}. As a result, the mass–radius relations of these two classes of compact objects differ qualitatively, providing a potential observational avenue for distinguishing their internal compositions. The identification of unusually compact low-mass objects, such as the central compact object in the supernova remnant HESS J1731–347, has therefore attracted considerable attention \cite{Doroshenko2016,Beznogov2020}.

Recent years have witnessed a rapid expansion of multimessenger constraints on compact stars. Precise mass measurements of heavy pulsars with masses close to 2 M$_\odot$ \cite{Demorest2010,Antoniadis2013,Cromartie2020}, radius measurements from NICER and XMM-Newton pulse-profile modeling \cite{Miller2019,Riley2019,Miller2021,Riley2021}, and tidal deformability constraints from the binary neutron star merger GW170817 \cite{Abbott2017,Abbott2018} together provide unprecedented information on the dense-matter EOS. In addition, the detection of the massive compact object in GW190814 \cite{Abbott2020GW190814} has raised renewed interest in the maximum mass of compact stars and the possible role of exotic matter at extreme densities.
At the same time, these observations have highlighted the importance of statistical inference and prior assumptions. Bayesian methods have become a standard tool for combining heterogeneous observational data and quantifying uncertainties in EOS modeling \cite{Steiner2010,Steiner2013,De2018_PRL121-091102,Xie2019_ApJ883-174,Xie2023_NST34-91,Xie2024_PRD110-043025}. Recent Bayesian analyses have explored both hadronic and quark matter EOSs under multimessenger constraints, demonstrating that different prior choices can lead to markedly different conclusions regarding the stiffness of dense matter and the viability of quark stars \cite{Lattimer2014_EPJA50-40,Xie2020_ApJ899-4,Annala2020,Raaijmakers2021}.

Despite this progress, several fundamental questions remain unresolved. To what extent do current observations constrain the microscopic structure of quark matter, as opposed to merely its overall stiffness? Can different quark phases be distinguished on the basis of macroscopic observables? And how robust are interpretations of extreme-massive objects, such as the secondary component of GW190814, against changes in prior assumptions?
In this work, we address these questions through a comprehensive Bayesian analysis of quark star models under current multimessenger constraints. Quark matter is described within an interacting MIT bag–model framework that consistently incorporates different color-superconducting phases. By systematically varying prior assumptions and assessing the impact of including or excluding the GW190814 secondary mass~\footnote{Note that recently, analyses on postmerger evolution of GW170817 indicates an upper bound on maximum mass of quark stars around $2.35 M_\odot$~\cite{Zhou:2024syq}, which however assumes a binary quark star merger. Thus, this is not a universal bound considering two-family scenario where the binary is a massive quark star and a less massive neutron star.}, we investigate how observational data constrain both microscopic quark-matter parameters and macroscopic stellar properties. Particular attention is paid to identifying which aspects of quark matter are robustly constrained by data and which remain dominated by prior choices.

The paper is organized as follows. In Sec. ~\ref{sec:the}, we introduce the quark star EOS models and outline the Bayesian inference framework employed in this study. Section ~\ref{sec:results} presents the resulting constraints on EOS parameters and stellar observables, with a detailed discussion of prior dependence and the role of extreme-mass observations. Our conclusions and implications for future multimessenger studies are summarized in Sec. ~\ref{sec:con}.

\section{\label{sec:the}EQUATIONS OF STATE AND BAYESIAN FRAMEWORK}
In this section, we describe the equation-of-state models adopted for quark stars together with the Bayesian inference framework used to confront them with multimessenger observations. Particular emphasis is placed on clarifying the physical role of the model parameters and on identifying which aspects of the EOS are expected to be constrained by current data.

\subsection{\label{sec:4dqsModel} Quark matter equation of state}
Various theoretical models have been proposed to describe the properties of quark matter in the literature, e.g., perturbation model~\cite{Fraga2014_ApJ781-L25, Kurkela2014_ApJ789-127, Xu2015_PRD92-025025, Xia2017_NPB916-669, Xia2019_PRD99-103017}, linear sigma model~\cite{Holdom2018_PRL120-222001}, MIT bag model~\cite{Zhou2018_PRD97-083015, Miao2021_ApJ917-L22}, Dyson-Swinger equations~\cite{Roberts1994_PPNP33-477, Alkofer2001_PR353-281}, equivparticle model~\cite{Peng2008_PRC77-065807, Xia2014_PRD89-105027}, quasiparticle model~\cite{Pisarski1989_NPA498-423, Schertler1997_JPG23-2051, Schertler1997_NPA616-659}, and Nambu–Jona-Lasinio model~\cite{Buball2005_PR407-205, Gholami2025_PRD111-103034}. Among these, the most widely used is the MIT bag model due to its simplicity and effectiveness in capturing certain essential features of QCD~\cite{Zhou2018_PRD97-083015, Miao2021_ApJ917-L22}. In the present work,
we model cold, deconfined quark matter within an interacting MIT bag–model framework, which provides a phenomenological but flexible description of quark matter at densities relevant for compact stars as presented in Refs.~\cite{Pereira2018_ApJ860-12, Zhang2021_PRD103-063018}.
The QM EOS can be expressed in terms of the pressure $P$ and energy density $\varepsilon$ as
\begin{equation}\label{eq:eos}
P=\frac{1}{3}(\varepsilon-4B)+ \frac{4\lambda^2}{9\pi^2}\left(-1+{\mathrm{sign}(\lambda)}\sqrt{1+3\pi^2 \frac{(\varepsilon-B)}{\lambda^2}}\right)
\end{equation}
with
\begin{equation}\label{eq:lam}
\lambda=\frac{c_2 \Delta^2-c_3 m_s^2}{\sqrt{c_1 a_4}},
\end{equation}
where the parameter set ($c_1$, $c_2$, $c_3$) take the values (1.86, 1, 0) for the 2SC phase; (3, 1, 0.75) for the 2SC+s phase; and (3, 3, 0.75) for the CFL phase, respectively~\cite{Zhang2021_PRD103-063018}.
The model is characterized by four parameters: $B$, $\Delta$, $m_s$, and $a_4$. Accordingly, we term it the four-parameter quark star model.
The effective bag constant, $B$, which reflects the contributions of the QCD vacuum, is typically treated as a phenomenological parameter. Meanwhile, the parameter $a_4$ represents the QCD corrections arising from gluon-mediated interactions between quarks. It ranges from 0 to 1, where $a_4$ = 1 indicates the absence of QCD corrections, and $a_4$ = 0 signifies extremely strong QCD corrections~\cite{Fraga2001_PRD63_121702,Alford2005_ApJ629-969,Bhattacharyya2016_MNRAS457-3101,Li2017_ApJ844-41,Zhang2021_PRD103-063018,Miao2021_ApJ917-L22}. The term involving the gap parameter $\Delta$ accounts for the effects of color superconductivity with $\Delta$ denoting the energy gap resulting from the quark pairing. $m_s$ is the strange quark mass having a range of 90 to 100 MeV~\cite{Olive2014_CPC38-090001}. This parameterized interacting QM EOS has been widely applied in recent years~\cite{Zhang:2021fla,Zhang:2021iah,Blaschke:2022egm,Koliogiannis:2022uim,Zhou:2024syq,Zhang:2022pse,Wang2024_JCAP2024-038,banerjee2024properties,Gammon:2023uss,gammon2025charged,Yuan:2023dxl,Pretel:2023nlr,Oikonomou:2023otn,Pretel:2024pem} thus serves as a representative working QM model.

The above EOS expression (\ref{eq:eos}) is derived from thermodynamic relations:
\be
P=-\Omega, \,\, n_{q}=-\frac{\partial\Omega}{\partial \mu},\,\, n_{e}=-\frac{\partial\Omega}{\partial \mu_e},\,\,   \varepsilon=\Omega+n_q \mu+n_e \mu_e ,
\label{eq:thermo}
\ee
Here $\Omega$ denotes the free energy of the superconducting quark matter and can be written as~\cite{Fraga2001_PRD63_121702,Alford2005_ApJ629-969,Weissenborn2011_ApJ740-L14}:
\begin{equation}\begin{aligned}
\Omega=&-\frac{c_1 a_4}{4\pi^2}\mu^4-\frac{\mu_{e}^4}{12 \pi^2}- \frac{c_2\Delta^2-c_3m_s^2}{\pi^2}  \mu^2+B,
\label{eq:omega}
\end{aligned}\end{equation}
where $\mu_e$ is the electron chemical potential, and the average quark chemical potential $\mu=(\mu_u+2\mu_d)/3$ for 2SC phase, and $\mu=(\mu_u+\mu_d+\mu_s)/3$ for 2SC+s and CFL phase with $\mu_u$, $\mu_d$ and $\mu_s$ being the chemical potentials for $u$, $d$ and $s$ quark, respectively. The expressions (\ref{eq:thermo}) can be thus rewritten as:
\bea
n_q&=&\frac{c_1a_4}{\pi^2}\mu^3 + \frac{2\lambda\sqrt{c_1 a_4}}{\pi^2}\mu, \quad  n_e=\frac{\mu_e^3}{3\pi^2},
\label{eq:rho_qe}
\eea
\bea
\varepsilon&=&\frac{3c_1 a_4}{4\pi^2}\mu^4+\frac{\mu_{e}^4}{4 \pi^2}+B+ \frac{ \lambda\sqrt{c_1a_4} }{\pi^2}  \mu^2.
\label{eq:epson}
\eea
The energy per baryon $E/A=3\mu$ under the condition of $P=0$ indicates the minimum energy per baryon~\cite{Zhang2021_PRD103-063018}, i.e.,
\bea
\frac{E}{A}
=\frac{3\sqrt{2}\pi}{(c_1 a_4)^{1/4}}\frac{B^{1/4}}{\sqrt{\sqrt{4\overline{\lambda}+\pi^2}+{2\text{sign}(\lambda)}\sqrt{\overline{\lambda}}}},
\label{eq:epa}
\eea
with $\overline{\lambda}=\lambda^2/4B$. For quark stars to exist stably, the  matter at their surfaces should be more stable than nuclear matter~\cite{Bodmer1971_PRD4-1601, Witten1984_PRD30-272, Terazaw1989_JPSJ58-3555}, i.e.,
\bea
\frac{E}{A}<\frac{M({^{56}\mathrm{Fe}})}{56}.
\label{eq:epa_max}
\eea
Meanwhile, since finite nuclei do not decay into $ud$QM nuggets, they should be unstable at $A\lesssim 300$ compared with that of finite nuclei. This condition provide additional constraints on $ud$QM and in particular the surface tension $\sigma$. If we demand the heaviest $\beta$-stable nucleus $^{266}$Hs to be more stable than $ud$QM nuggets, the energy per baryon of $ud$QM in 2SC should fulfill the following stability constraint, i.e.,
\bea
\frac{M({^{266}\mathrm{Hs}})}{266} < \frac{E}{A} +\sigma\left(\frac{An_0^{2}}{36\pi}\right)^{-1/3},
\label{eq:epa_266Hs}
\eea
where the left hand side represents the energy per baryon for ${^{266}\mathrm{Hs}}$ and on the right hand side a liquid-drop formula is employed with $A$=266 and $n_0=n_q/3$ the saturation density of $ud$QM at $\mu = E/3A$ fixed by Eq.~(\ref{eq:epa}). This indicates a minimum value for $\sigma$ at fixed bag model parameters $B$, $\Delta$, and $a_4$ with $c_1=1.86$, $c_2=1$, and $c_3=0$ for the 2SC phase, i.e.,
\bea
\sigma_{\rm{min}}=\left(\frac{An_0^{2}}{36\pi}\right)^{1/3}(931.74\ \mathrm{MeV}-\frac{E}{A}).
\label{eq:sigma_min}
\eea

\subsection{\label{sec:3dqsModel}  Effective parameter reduction and physical interpretation}
The full quark matter model formally depends on several parameters; however, they could be correlated in microscopic physics, and not all of them are independently constrained by macroscopic observables. In particular, we find that variations in the strange quark mass and the pairing gap ($\Delta$) lead to largely degenerate effects on the pressure–energy-density relation once the effective bag constant is fixed.
To make this redundancy explicit, we recast the quark matter EOS in terms of a reduced set of effective parameters that directly control the macroscopic stiffness of the EOS. The dominant parameter is the effective bag constant $B$, which determines both the onset density of deconfinement and the maximum mass of quark stars. Additional parameters enter only through subleading corrections and primarily affect the EOS at very high densities.
This effective parameter reduction allows us to explore a broad class of quark matter models while avoiding unnecessary dimensionality in the Bayesian inference. Importantly, it also clarifies the physical interpretation of the results: constraints obtained from current observations primarily reflect limits on the overall stiffness of quark matter, rather than on the detailed microscopic structure or pairing pattern.

We can convert Eq. (\ref{eq:eos}) into a more general form without the explicit sign$(\lambda)$ factor by introducing the following dimensionless rescaling:
\begin{equation}
g=\frac{\lambda}{2\sqrt{B}}=\text{sign}(\lambda)\sqrt{\bar{\lambda}}
\label{scaling_prho}
\end{equation}
so that Eq. (\ref{eq:eos}) reduces to:
\begin{equation}
\frac{P}{4B}=\frac{1}{3}(\frac{\rho}{4B}-1)+\frac{4}{9\pi^2}\left(-g^2+g\sqrt{g^2+3\pi^2(\frac{\rho}{4B}-\frac{1}{4})}\right)
\label{eos_3para}
\end{equation}
and Eq.(\ref{eq:epa}) can be rewritten as:
\be
\frac{E}{A}=\frac{3\sqrt{2}\pi}{(c_1 a_4)^{1/4}}\frac{B^{1/4}}{\sqrt{\sqrt{4g^2+\pi^2}+2g}}.
\label{eq:epa2}
\ee
In the equation above, there are four parameters, i.e., $B$, $g$, $a_4$ and $c_1$. 
However, the parameter $c_1$ varies only slightly among the quark matter phases 2SC, 2SC+s, and CFL. Therefore, we neglect the $c_1$ parameter in this study. This is justified by our calculations, as shown in Fig. \ref{fig:3para-pr2}, where the posterior distributions of $B$, $a_4$, $g$, and $E/A$ obtained with different $c_1$ values show negligible differences. Based on this simplification, we refer to the resulting model as the three-parameter quark star model.

As illustrated by the posterior distribution of the energy per baryon ($E/A$) in Fig. \ref{fig:3para-pr2}, and considering current neutron star observations, it is particularly interesting to note that under the prior ranges adopted for the four parameters of the quark star EOS model, the resulting $E/A$ of quark matter consistently satisfies the stability condition $E/A \leq$ 930 MeV. Consequently, this stability condition does not need to be imposed as an additional constraint in our Bayesian analysis. This reduces the number of free parameters in the quark star EOS to just two: $B$ and $g$. Thus, we refer to this final version as the two-parameter quark star model.

\subsection{\label{sec:mr}Stellar structure and observable quantities}
Given an EOS, the structure of nonrotating compact stars is obtained by solving the Tolman–Oppenheimer–Volkoff (TOV) equations~\cite{Tolman1934_PNAS20-169, Oppenheimer1939_PR55-374}, i.e.,
\begin{eqnarray}
&&\frac{\mbox{d}P}{\mbox{d}r} = -\frac{G M \varepsilon}{r^2}   \frac{(1+P/\varepsilon)(1+4\pi r^3 P/M)} {1-2G M/r},  \label{eq:TOV}\\
&&\frac{\mbox{d}M}{\mbox{d}r} = 4\pi \varepsilon r^2. \label{eq:m_star}
\end{eqnarray}
The gravitational constant is given by $G = 6.707 \times 10^{-45}\ \mathrm{MeV}^{-2}$.
The dimensionless tidal deformability, $\Lambda$, is calculated using the following expression~\cite{Fattoyev2013_PRC87-015806, Malik2018_PRC98-035804, Hinderer2010_PRD81-123016, Hinderer2008_ApJ677-1216}:
\begin{equation}\label{eq:Lambda}
  \Lambda = \frac{2}{3} k_2 \left( \frac{R}{M} \right)^{5},
\end{equation}
where $R$ and $M$ are the radius and mass of the quark star, respectively. The second Love number, $k_2$, is determined by the equation of state and is governed by a set of coupled equations, specifically Eqs. (3)-(6) in Ref.~\cite{Malik2018_PRC98-035804}, which are solved alongside the TOV equations with appropriate boundary conditions. Note that a special boundary condition at $r=R$ is adopted for QSs due to their large surface densities~\cite{Damour2009_PRD80-084035, Postnikov2010_PRD82-024016}.

\subsection{\label{sec:bayes}Bayesian inference framework}
We employ Bayesian inference to constrain EOS parameters using multimessenger observations. For a set of model parameters $\bm{\theta}$, Bayes’ theorem gives the posterior probability distribution
\begin{align}
p(\bm{\theta} \,|\, &\bm{D}, {\cal M})
\propto
p(\bm{\theta} \,|\, {\cal M})\times p_{\textnormal{filter}} \times p_{\rm{mass,max}} \nonumber\\
& \times p(M_{\textnormal{max}} \,|\, \bm{D}_{\textnormal{GW190814}}) \nonumber\\
& \times p(M, R \,|\, \bm{D}_{\textnormal{HESS}}) \nonumber\\
& \times \prod_{i} p(M_i, R_i \,|\, \bm{D}_{\textnormal{NICER},i}) \,,
\label{eq:post}
\end{align}
where $\bm{\theta}$= $B$, $\Delta$, $m_s$, and $a_4$ for the four-parameter quark star model with their prior ranges listed in Table \ref{tab-prior}, and $\bm{\theta}$= $B$, $\sqrt[4]{\bar{\lambda}}$ and $a_4$ for the three-parameter quark star model with the prior ranges of 0 to 200 MeV/fm$^3$, 0 to 100 and 0 to 1, respectively. $\bm{\theta}$= $B$ and $\sqrt[4]{\bar{\lambda}}$  for the two-parameter quark star model with the prior ranges of 0 to 200 MeV/fm$^3$ and 0 to 100, respectively. Note that the variation of parameters could lead to different quark phases such as gapless phases, crystalline phases, and meson condensations~\cite{Alford2008_RMP80-1455, Lai2009_MMRAS398-L31}, which are nonetheless neglected here.
$p(M_i, R_i \,|\, \bm{D}_{\textnormal{NICER},i})$ is the NICER likelihood function with $\bm{D}_{\rm{NICER}}$ being the observational data from the NICER collaboration on the PSR J0030+0451, PSR J0740+6620 and PSR 0437-4715 as listed in Table \ref{tab-data}. $\bm{D}_{\rm{HESS}}$ denotes the lower-mass object HESS J1731-347 and $\bm{D}_{\rm{GW190814}}$ represents the second component of GW190814.

Based on the posterior distribution defined in Eq. (\ref{eq:post}), we generate samples from the uniform prior, $p(\bm{\theta} \,|\, {\cal M})$. Note that the uniform prior adopted for the four-parameter model will lead to a nonuniform prior for the three-parameter model according to Eq. (\ref{scaling_lam}), while here we adopt a uniform prior for the three-parameter model instead and could lead to a different  posterior distribution. Similar scenarios are expected among four/three/two-parameter models. For each sampled set of parameters, the corresponding stellar properties, such as mass ($M$), radius ($R$), and tidal deformability ($\Lambda$), are computed. The likelihood function is then evaluated by comparing these predicted values to the astrophysical observational constraints, which is achieved by applying a kernel density estimation to the posterior distributions obtained from independent astrophysical analyses of neutron star observables.

The component $p_{\rm{filter}}$ acts as a filter to select EOS parameter sets that satisfy the following conditions: (1) the thermodynamic stability condition, i.e., $\mbox{d}P/\mbox{d}\varepsilon \geq 0$; (2) the causality condition, ensuring that the speed of sound is always less than the speed of light at all densities; (3) the energy per baryon for quark matter at $P=0$ is not greater than the one for the most stable nucleus $^{56}$Fe, i.e., $E/A \leq 930$ MeV; (4) the energy per baryon of $ud$QM nuggets must exceed 931.742 MeV at $A=266$, i.e., fulfilling Eq.~(\ref{eq:epa_266Hs}); (5) the tidal deformability for compact stars with 1.4 $\mathrm{M}_{\odot}$ should not be greater than 800, i.e., $\Lambda_{1.4} \leq$ 800.
The term $p_{\textnormal{mass,max}}$ ensures that the EOS is sufficiently stiff to support the observational maximum mass of neutron stars, $M_{\textnormal{max}}$. In this analysis, we adopt a value of $M_{\textnormal{max}} = 1.97 M_{\odot}$.
To simulate the posterior probability density function (PDF) of the model parameters, we employ a Markov-Chain Monte Carlo (MCMC) method using the Metropolis-Hastings algorithm. This method allows us to sample the parameter space and derive the PDFs for individual parameters and their pairwise correlations by integrating over the other parameters. 

During the MCMC sampling, it is essential to discard the initial samples from the burn-in period, as the algorithm does not yet sample from the equilibrium distribution at the start. Based on our previous work, we find that 40,000 burn-in steps are sufficient to reach equilibrium when the model contains six parameters~\cite{Xie2019_ApJ883-174, Xie2020_ApJ899-4, Xie2023_NST34-91}. Given that the current analysis involves up to four parameters, this burn-in step count is more than adequate. Therefore, we discard the first 40,000 steps and utilize the remaining one million steps to compute the posterior PDFs of the four parameters in this analysis.

\begin{table}[ht]
\centering
\caption{Prior ranges of the parameters used in this work. The five independent parameters employed in this study utilize two types of prior ranges. The parameters $a_4$, $m_s$, and $\sigma$ adopt fixed prior ranges, while the parameters $B$ and $\Delta$ are assigned two different sets of ranges as marked by Prior-1 and Prior-2.
}\label{tab-prior}
 \begin{tabular}{lcccccccc}
  \hline\hline
   Parameters& Prior-1  & Prior-2 \\
    \hline 
$B$(MeV/fm$^{3}$) & (1,100) & (1,200) \\
$\Delta$(MeV) & (0,100) & (0,1000) \\
$a_4$  & (0.1,1) & (0.1,1) \\
$m_s$(MeV) & (90,100) & (90,100) \\
$\sigma$(MeV/fm$^{2}$)& (0,300) &(0,300) \\
 \hline\hline
 \end{tabular}
\end{table}

\begin{table}[htbp]
\centering
\caption{Data for a compact star's mass and radius  (68\% CI) used in the present work.}\label{tab-data}
 \begin{tabular}{lccccccc}
  \hline\hline
  & Mass($\mathrm{M}_{\odot}$)&Radius $R$ (km)  & Source and Reference \\
    \hline 
 &$1.34_{-0.16}^{+0.15}$ &$12.71_{-1.19}^{+1.14}$ &PSR J0030+0451~\cite{Riley2019_ApJ887-L21} \\
&$1.48_{-0.037}^{+0.037}$ &$11.36_{-0.63}^{+0.95}$ &PSR J0437-4715~\cite{Choudhury2024_ApJ971-L20} \\
&$2.072_{-0.0066}^{+0.067}$ &$12.39_{-0.98}^{+1.30}$ &PSR J0740+6620~\cite{Riley2021_ApJ918-L27} \\
&$0.77_{-0.17}^{+0.20}$ &$10.4_{-0.78}^{+0.86}$&HESS J1731-347~\cite{Doroshenko2022_NA6-1444}\\
&$2.59_{-0.09}^{+0.08}$ &-- &GW190814~\cite{Abbott2020_Astrophys.J.896-L44}\\
  \hline\hline
\end{tabular}
\end{table}

\section{\label{sec:results}RESULTS AND DISCUSSION}
In this section, we present the Bayesian constraints on quark star equations of state
inferred from current multimessenger observations.
We first examine how the observational data constrain macroscopic stellar properties,
including mass--radius relations, the posterior distributions for radii, tidal deformabilities,
maximum masses, maximum radius, central energy density and central pressure of quark stars.
We then analyze the posterior distributions of the microscopic parameters
characterizing quark matter and their mutual correlations.

A central aspect of this section is a systematic comparison between two prior
prescriptions and between analyses performed with and without including the
GW190814 secondary component.
This allows us to assess the role of extreme-mass observations in shaping the inferred
high-density behavior of quark matter and to identify which conclusions are robust
against prior choices and which are primarily driven by the inclusion of GW190814.
Throughout this section, we emphasize the physical origin of the observed prior
dependence in the Bayesian inference.

\subsection{\label{sec:mass-rad}Global mass–radius properties of quark stars}
\begin{table*}[htbp]
\centering
\caption{Summary of posterior estimates for the derived quantities of quark stars (radius and tidal deformability for 1.4 and 2.0 solar mass stars, maximum mass and the corresponding radius), central pressure and central energy density for the CFL, 2SC+s and 2SC phases. Values are inferred using the combined datasets listed in Table~\ref{tab-data} with considering the constraints of the second component of GW190814, and reported as median estimates with 90\% credible intervals. Their prior ranges are indicated in Table~\ref{tab-prior}.}\label{tab-6parampv-gw}
\begin{tabular}{lccccccc}
  \hline\hline
   Parameters&Prior-1  &Prior-2 \\
    \hline 
   &2SC, 2SC+s, CFL&2SC, 2SC+s, CFL \\
   \hline 
$R_{1.4}$(km) & $11.9_{-0.2}^{+0.0}$,$11.9_{-0.2}^{+0.0}$,$11.8_{-0.2}^{+0.1}$
                                  &$11.3_{-0.2}^{+0.4}$,	$11.3_{-0.2}^{+0.5}$,	$11.3_{-0.2}^{+0.4}$\\
$R_{2.0}$(km)& $12.7_{-0.3}^{+0.0}$,	$12.7_{-0.3}^{+0.0}$,	$12.8_{-0.3}^{+0.0}$
                                 &$12.6_{-0.2}^{+0.5}$,	$12.8_{-0.4}^{+0.3}$,	$12.5_{-0.2}^{+0.5}$ \\
$\Lambda_{1.4}$ &	$796.0_{-82.3}^{+4.0}$,	$796.5_{-82.3}^{+3.5}$,	$794.4_{-82.6}^{+5.6}$
                                  &$690.8_{-85.0}^{+106.5}$,	$690.8_{-83.2}^{+107.4}$,	$689.9_{-107.7}^{+105.0}$ \\
$\Lambda_{2.0}$ &$104.6_{-19.8}^{+1.6}$,	$90.4_{-6.3}^{+15.7}$,	$108.7_{-20.1}^{+2.4}$
                                   &$121.2_{-21.7}^{+25.9}$,	$121.5_{-21.0}^{+25.8}$,	$120.5_{-23.5}^{+25.8}$  \\
$M_{\texttt{max}}$(M$_{\odot}$) & $2.2_{-0.0}^{+0.0}$,	$2.3_{-0.1}^{+0.0}$,	$2.3_{-0.1}^{+0.0}$
                                  &$3.2_{-0.1}^{+0.1}$,	$3.2_{-0.1}^{+0.2}$,	$3.2_{-0.1}^{+0.2}$\\
$R_{\texttt{max}}$(km)& $12.3_{-0.3}^{+0.0}$,	$12.3_{-0.3}^{+0.0}$,	$12.3_{-0.3}^{+0.1}$
                                 &$13.6_{-0.6}^{+0.4}$,	$13.6_{-0.5}^{+0.4}$,$13.4_{-0.4}^{+0.6}$\\
$\varepsilon_{c}$ (MeV/fm$^{3}$)&$887.5_{-8.4}^{+46.5}$,	$889.5_{-7.3}^{+47.4}$,	$874.6_{-10.6}^{+47.6}$
                              &$714.5_{-39.9}^{+62.1}$,	$714.9_{-39.3}^{+61.4}$,	$733.7_{-61.6}^{+47.5}$\\
$P_{c}$(MeV/fm$^{3}$) &$242.6_{-6.1}^{+11.7}$,	$239.9_{-3.4}^{+12.2}$,	$245.4_{-6.1}^{+14.8}$
                     &$466.3_{-79.6}^{+45.8}$,	$454.7_{-75.7}^{+53.9}$,	$476.2_{-64.0}^{+47.2}$ \\
 \hline
 \end{tabular}
\end{table*}
\begin{table*}[htbp]
\centering
\caption{Summary of posterior estimates for the derived quantities of quark stars (radius and tidal deformability for 1.4 and 2.0 solar mass stars, maximum mass and the corresponding radius), central pressure and central energy density for the CFL, 2SC+s and 2SC phases. Values are inferred using the combined datasets listed in Table~\ref{tab-data} without considering the constraints of the second component of GW190814, and reported as median estimates with 90\% credible intervals. Their prior ranges are indicated in Table~\ref{tab-prior}.}\label{tab-6parampv-nogw}
\begin{tabular}{lccccccc}
  \hline\hline
   Parameters&Prior-1  &Prior-2 \\
    \hline 
   &2SC, 2SC+s, CFL&2SC, 2SC+s, CFL \\
   \hline 
$R_{1.4}$(km) & $11.7_{-0.4}^{+0.4}$,$11.7_{-0.7}^{+0.4}$,$11.7_{-0.7}^{+0.4}$	
                                  &$11.6_{-0.9}^{+0.3}$,	$11.6_{-0.9}^{+0.3}$,	$11.5_{-0.9}^{+0.3}$	 \\
$R_{2.0}$(km)& $12.6_{-0.8}^{+0.6}$,	$12.5_{-0.9}^{+0.6}$,	$12.5_{-0.8}^{+0.6}$
                                 &$13.00^{+0.00}_{-0.70}$,$13.00^{+0.00}_{-0.70}$,$13.00^{+0.30}_{-0.40}$ \\
$\Lambda_{1.4}$ &$748.2_{-241.5}^{+51.0}$,	$700.8_{-274.1}^{+95.2}$,	$632.7_{-204.9}^{+159.5}$
                                  &$612.9_{-211.9}^{+171.2}$,	$613.6_{-214.9}^{+168.4}$,	$612.2_{-218.0}^{+175.2}$ \\
$\Lambda_{2.0}$ &$89.2_{-34.0}^{+25.4}$,	$72.5_{-41.7}^{+34.1}$,	$72.9_{-41.5}^{+34.9}$
                                   &$106.6_{-43.8}^{+48.2}$,	$106.9_{-46.6}^{+47.7}$,	$106.6_{-43.4}^{+48.2}$ \\
$M_{\texttt{max}}$(M$_{\odot}$) & $2.3_{-0.2}^{+0.1}$,	$2.2_{-0.2}^{+0.1}$,	$2.2_{-0.2}^{+0.1}$
                                  &$3.1_{-0.4}^{+0.3}$,	$3.1_{-0.9}^{+0.3}$,	$3.1_{-0.4}^{+0.3}$\\
$R_{\texttt{max}}$(km)& $12.0_{-0.7}^{+0.4}$,	$12.0_{-1.1}^{+0.4}$,	$11.9_{-0.9}^{+0.5}$	
                                 &$13.0_{-1.1}^{+1.1}$,	$13.0_{-1.2}^{+1.1}$,$13.4_{-1.4}^{+0.9}$\\
$\varepsilon_{c}$(MeV/fm$^{3}$) &$884.8_{-25.3}^{+145.8}$,	$954.5_{-82.3}^{+164.3}$,	$924.9_{-59.9}^{+177.6}$
                              &$751.5_{-96.9}^{+162.7}$,	$753.3_{-99.7}^{+183.1}$,	$736.0_{-85.9}^{+173.0}$\\
 $P_{c}$(MeV/fm$^{3}$)&	$250.8_{-19.3}^{+27.1}$,	$248.1_{-20.8}^{+44.7}$,	$248.4_{-19.6}^{+47.0}$
                     &$449.3_{-119.6}^{+124.5}$,	$443.0_{-145.3}^{+120.8}$,	$455.6_{-95.4}^{+147.6}$\\
 \hline
 \end{tabular}
\end{table*}
 \begin{figure*}[ht]
\begin{center}
  \includegraphics[width=13cm,height=10cm]{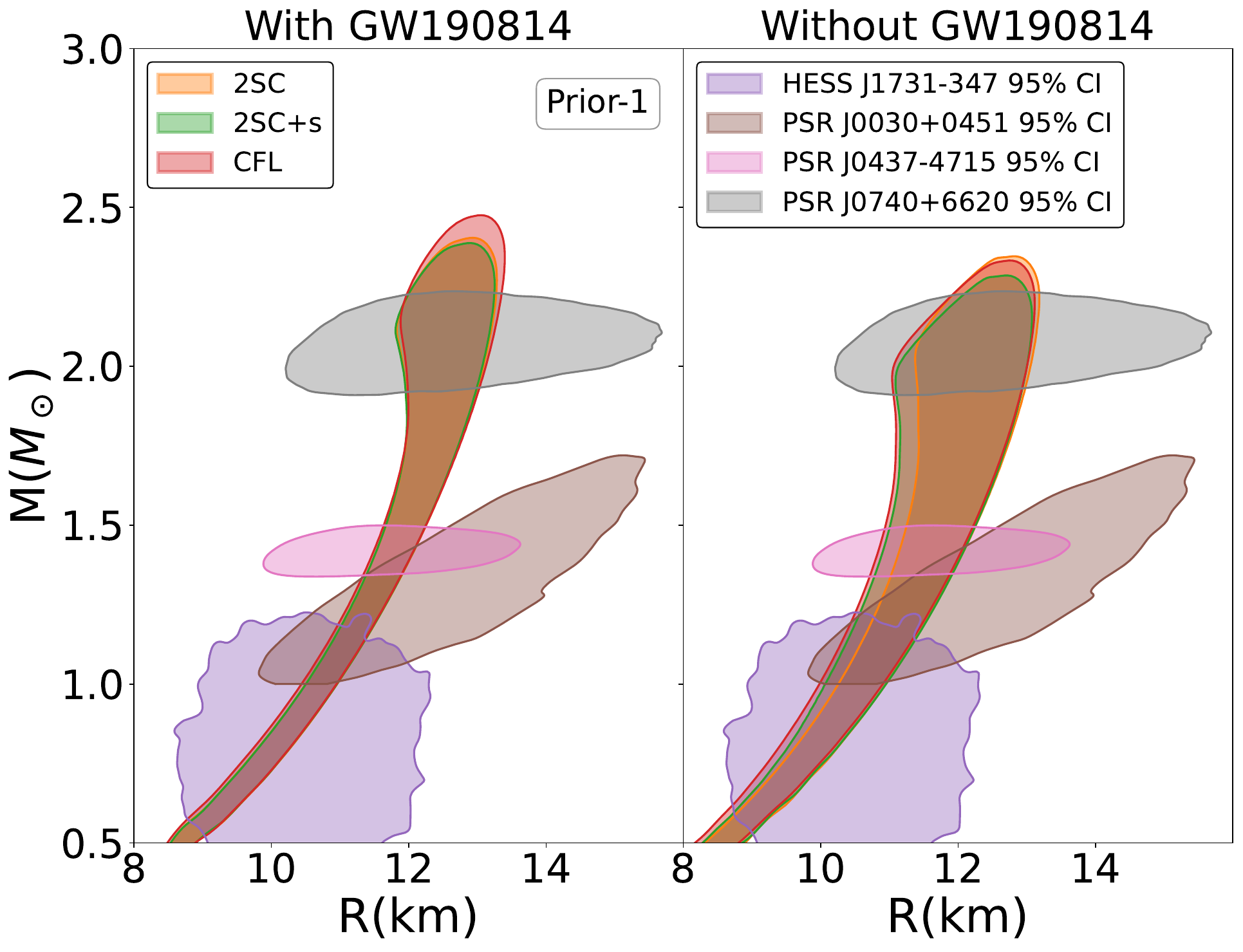}
  \caption{(Color online)
  Mass-radius constraints at $90\%$ credibility for quark stars in the CFL, 2SC+s, and 2SC phases. The calculations for quark stars are based on a four-parameter model with the ``Prior-1" parameter distributions. Observational constraints from NICER (PSRs J0030+0451~\cite{Miller2019_ApJ887-L24, Riley2019_ApJ887-L21}, J0740+6620~\cite{Miller2021_ApJ918-L28, Riley2021_ApJ918-L27}, J0437-4715\cite{Choudhury2024_ApJ971-L20}) and HESS J1731-347~\cite{Doroshenko2022_NA6-1444} are shown for comparison. The left and right panels correspond to cases with and without the inclusion of the secondary component of GW190814 as a constraint, respectively.}\label{fig:mr-pr1}
\end{center}
\end{figure*}
 \begin{figure*}[ht]
\begin{center}
  \includegraphics[width=13cm,height=10cm]{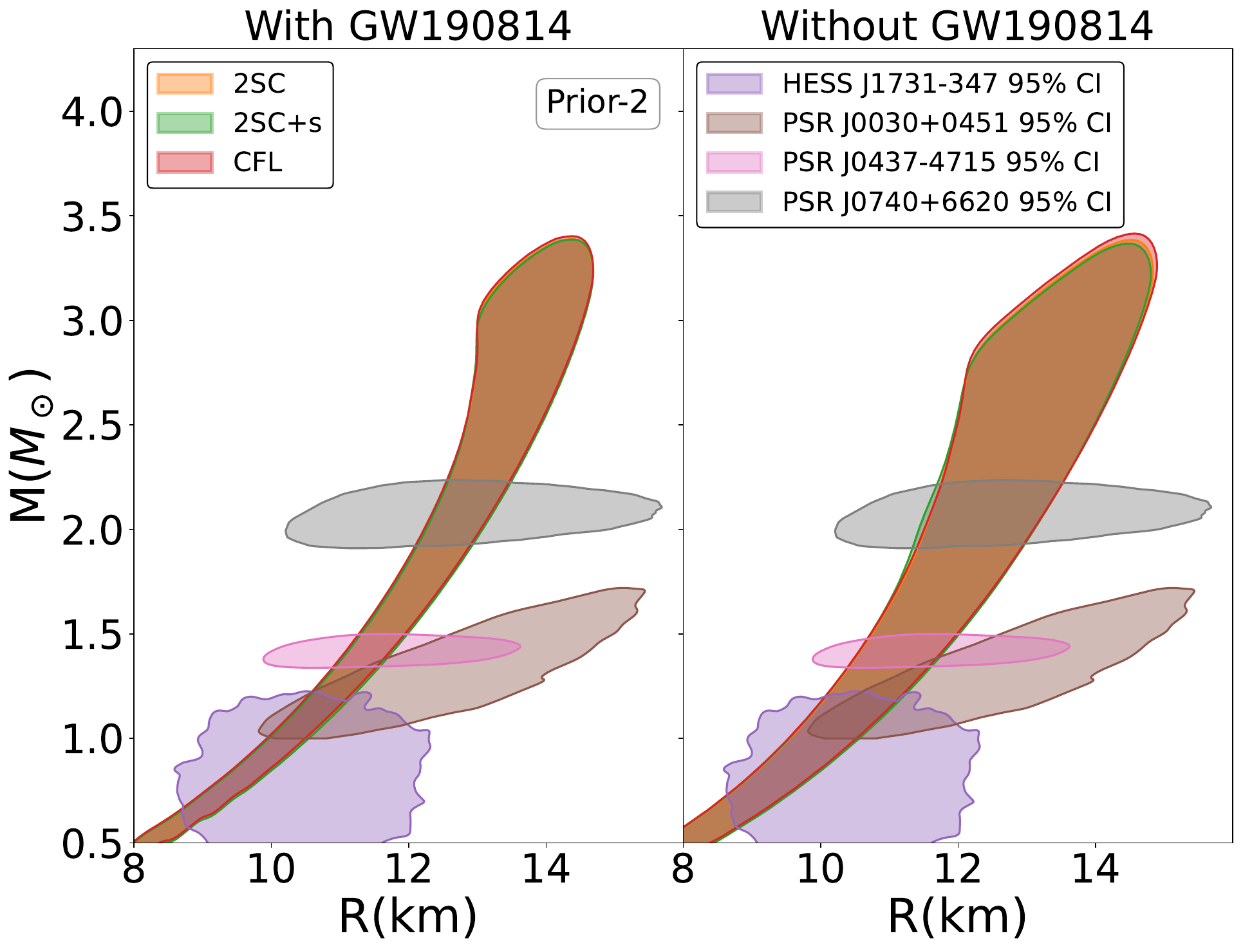}
  \caption{(Color online)
  As in Fig. \ref{fig:mr-pr1}, but showing the results based on the four-parameter model with the  ``Prior-2" parameter distributions.}\label{fig:mr-pr2}
\end{center}
\end{figure*}

We first examine the mass–radius relations of quark stars inferred from the Bayesian analysis based on the four-parameter quark matter model. Figures ~\ref{fig:mr-pr1} and ~\ref{fig:mr-pr2} show the 90\% credible regions for the CFL, 2SC+s, and 2SC phases under the two prior prescriptions. In all cases, the inferred mass–radius relations are fully consistent with current radius measurements from NICER for PSRs J0030+0451~\cite{Riley2019_ApJ887-L21}, J0740+6620~\cite{Riley2021_ApJ918-L27}, and J0437–4715~\cite{Choudhury2024_ApJ971-L20}, and they naturally accommodate the compact low-mass object HESS J1731–347~\cite{Doroshenko2022_NA6-1444}. It is important to emphasize that the existence of the ultra-low mass object HESS J1731-347 poses a severe challenge to standard hadronic equations of state, which typically exhibit a minimum mass limit higher than the observed value or struggle to reconcile such a small radius with nuclear constraints. In contrast, our results demonstrate that self-bound quark stars naturally accommodate this object without fine-tuning, as their mass-radius relation extends to arbitrarily low masses. This provides a distinct observational signature favoring the strange quark matter hypothesis in the low-mass regime.

A salient feature of these results is that the mass–radius curves corresponding to different quark phases are nearly indistinguishable within the inferred credibility regions. This insensitivity indicates that, once the bulk stiffness of quark matter is fixed by observational constraints, the detailed phase structure, whether 2SC, 2SC+s, or CFL, has only a marginal impact on the macroscopic stellar properties.
This degeneracy of different phases matches the expectation from the two-parameter QS model where the difference between different phases is subsumed in the parameter $\lambda$.

The influence of prior assumptions is most evident in the high-mass regime. Under the narrower Prior-1 setup, the maximum mass of quark stars is limited to approximately 2.3$\text{–}$2.5 M$_\odot$, whereas the broader Prior-2 allows significantly stiffer equations of state, extending the maximum mass up to $\sim3.4 M_\odot$. Consequently, only the Prior-2 configuration permits an interpretation of the GW190814 secondary component as a quark star, while the Prior-1 setup disfavors this possibility. Furthermore, incorporating the mass constraint from the GW190814 secondary component significantly narrows the allowable parameter space for the mass-radius relations of quark stars. Our analysis thus delineates the boundary of the parameter space: interpreting the secondary component of GW190814 as a quark star requires an extremely stiff equation of state, which is only accessible under the broader prior assumptions (Prior-2). If this object is indeed a compact star, it strongly disfavors soft quark matter models; conversely, if it is a black hole, the parameter space favored by standard pulsars (consistent with Prior-1) remains the most probable description of dense quark matter.

\subsection{\label{sec:macro-obser}Macroscopic observables from the four-parameter model}
 \begin{figure*}[ht]
\begin{center}
  \includegraphics[width=15cm,height=10cm]{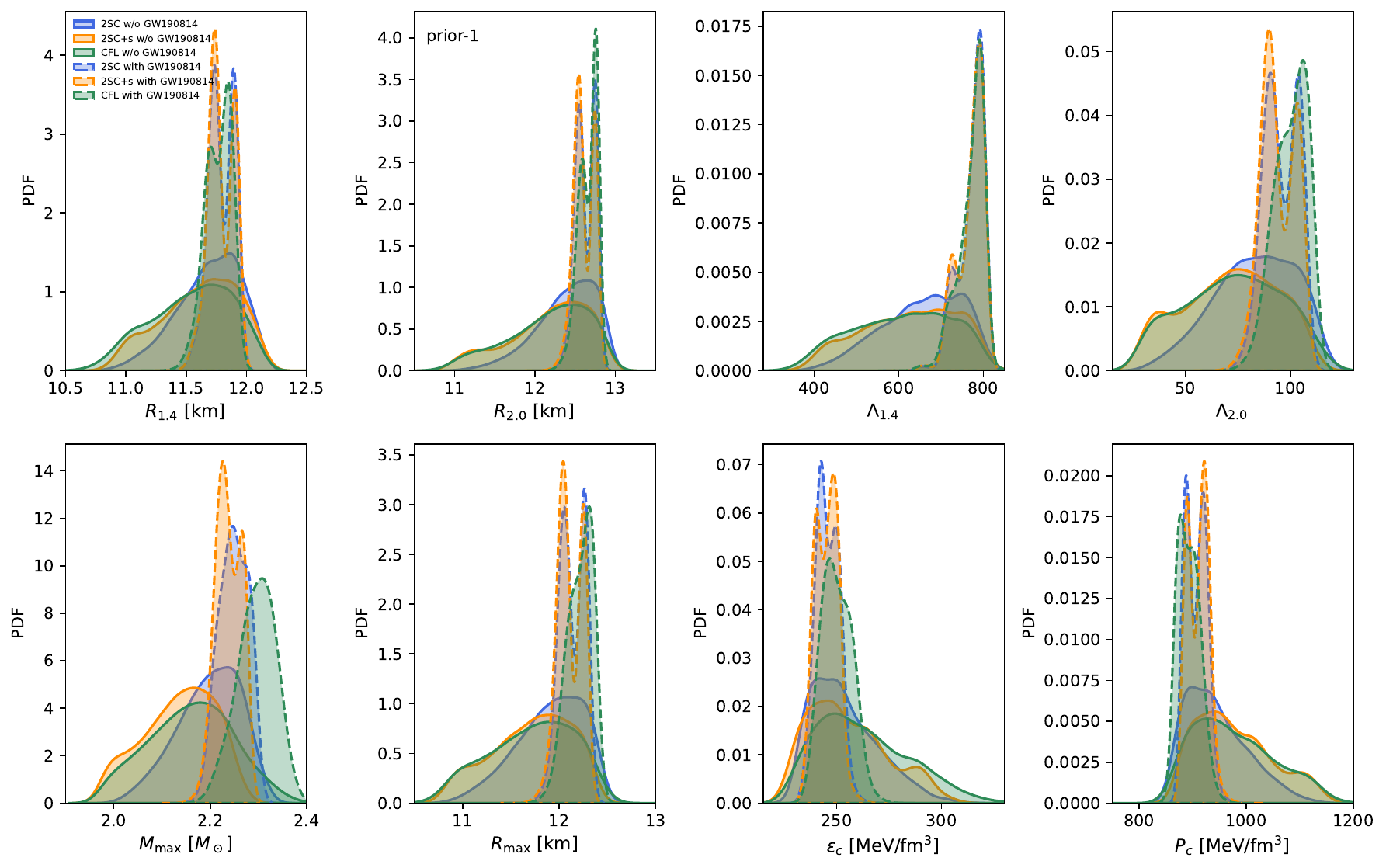}
  \caption{(Color online) Posterior probability distribution functions of the radius, tidal deformability corresponding to the quark stars with 1.4 and 2.0 solar masses, maximum mass and the corresponding radius, central squared speed of sound, central energy density and central pressure inferred from the Bayesian analysis of the data listed in Table \ref{tab-data} for the CFL, 2SC and 2SC+s phases.}\label{fig:8paraPDF-pr1}
\end{center}
\end{figure*}

 \begin{figure*}[ht]
\begin{center}
  \includegraphics[width=15cm,height=10cm]{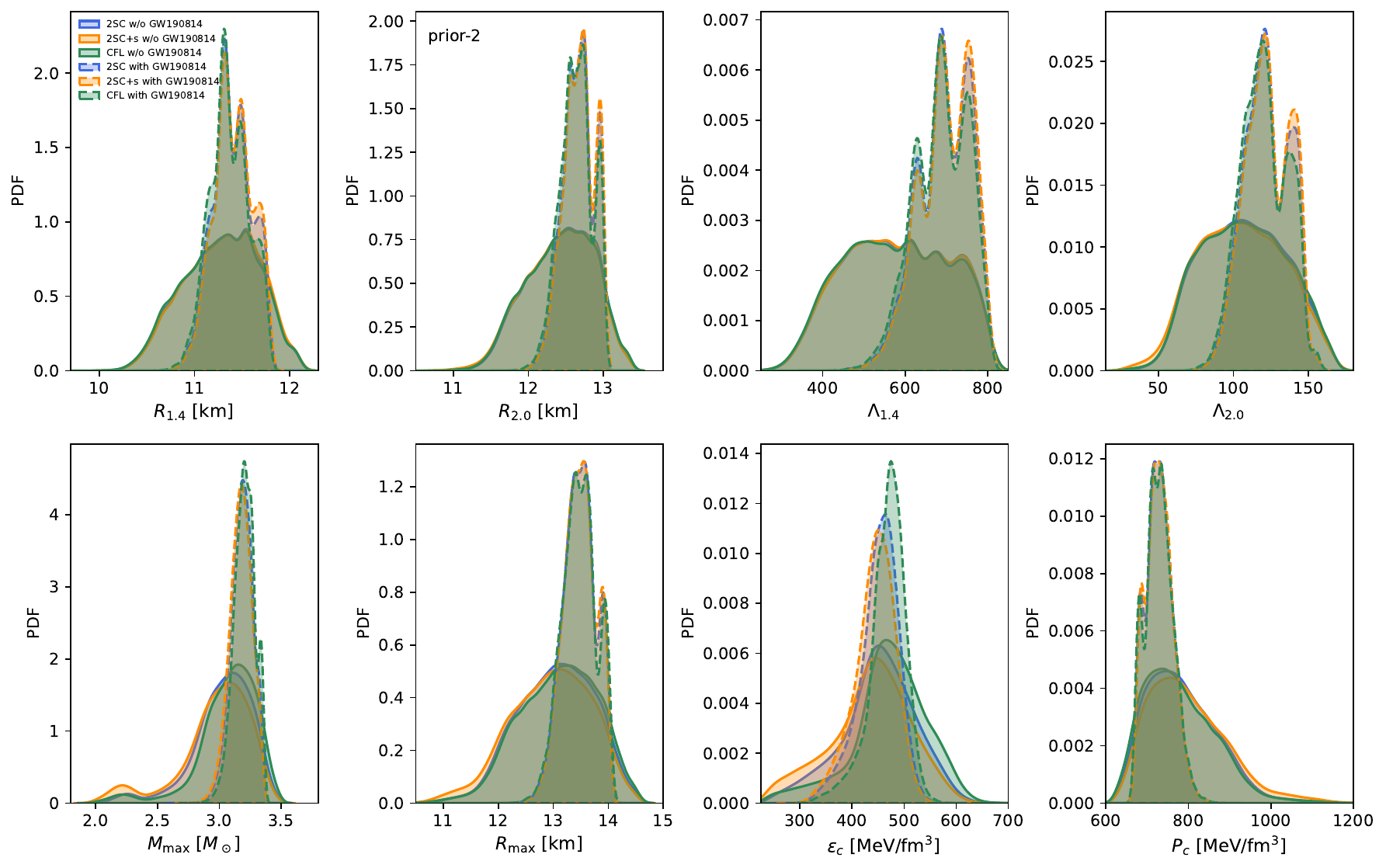}
  \caption{(Color online) Posterior probability distribution functions of the radius, tidal deformability corresponding to the quark stars with 1.4 and 2.0 solar masses, maximum mass and the corresponding radius, central squared speed of sound, central energy density and central pressure inferred from the Bayesian analysis of the data listed in Table \ref{tab-data} for the CFL, 2SC and 2SC+s phases.}\label{fig:8paraPDF-pr2}
\end{center}
\end{figure*}
The posterior distributions of macroscopic quark star properties, including the radius, tidal deformability of 1.4 $M_\odot$ and 2.0 $M_\odot$ stars, maximum mass and the corresponding radius, central energy density and central pressure, derived from the four-parameter model are shown in Figures ~\ref{fig:8paraPDF-pr1} and ~\ref{fig:8paraPDF-pr2} and summarized quantitatively in Tables ~\ref{tab-6parampv-gw} and ~\ref{tab-6parampv-nogw}.
For canonical masses of 1.4 $M_\odot$ and 2.0 $M_\odot$, the inferred radii are remarkably robust against variations in both the quark phase and the prior choice, typically clustering around $R_{1.4}\simeq 11.3\text{–}11.9~\mathrm{km}$ and $R_{2.0}\simeq 12.5\text{–}13.0~\mathrm{km}$.
In contrast, the tidal deformabilities exhibit a stronger sensitivity to prior assumptions. While $\Lambda_{1.4}$ generally decreases and $\Lambda_{2.0}$ increases when transitioning from Prior-1 to Prior-2, both quantities remain compatible with current gravitational-wave constraints. This behavior indicates that tidal deformability is more sensitive to the detailed stiffness of the EOS than the stellar radius, especially at higher masses.
The maximum mass and the corresponding radius show the strongest prior dependence. Inclusion of the GW190814 constraint significantly narrows the posterior distributions, particularly under the Prior-2 setup, demonstrating that extreme-mass observations provide powerful leverage on the high-density behavior of quark matter.

The extreme EOS in the cores of compact stars, representing the densest matter environments in the Universe, plays a pivotal role in understanding strong interaction physics and general relativistic effects. The pressure and energy density at the center of the quark star exhibit weak dependence on the inclusion of the GW190814 constraint but are sensitive to the choice of parameter prior. Employing the Prior-2 configuration results in an enhanced central pressure and a reduced central energy density compared to the case from Prior-1 setup.

\subsection{\label{sec:5paraPDF}Posterior constraints on microscopic quark matter parameters}
\begin{figure*}[ht]
\begin{center}
  \includegraphics[width=16cm,height=13cm]{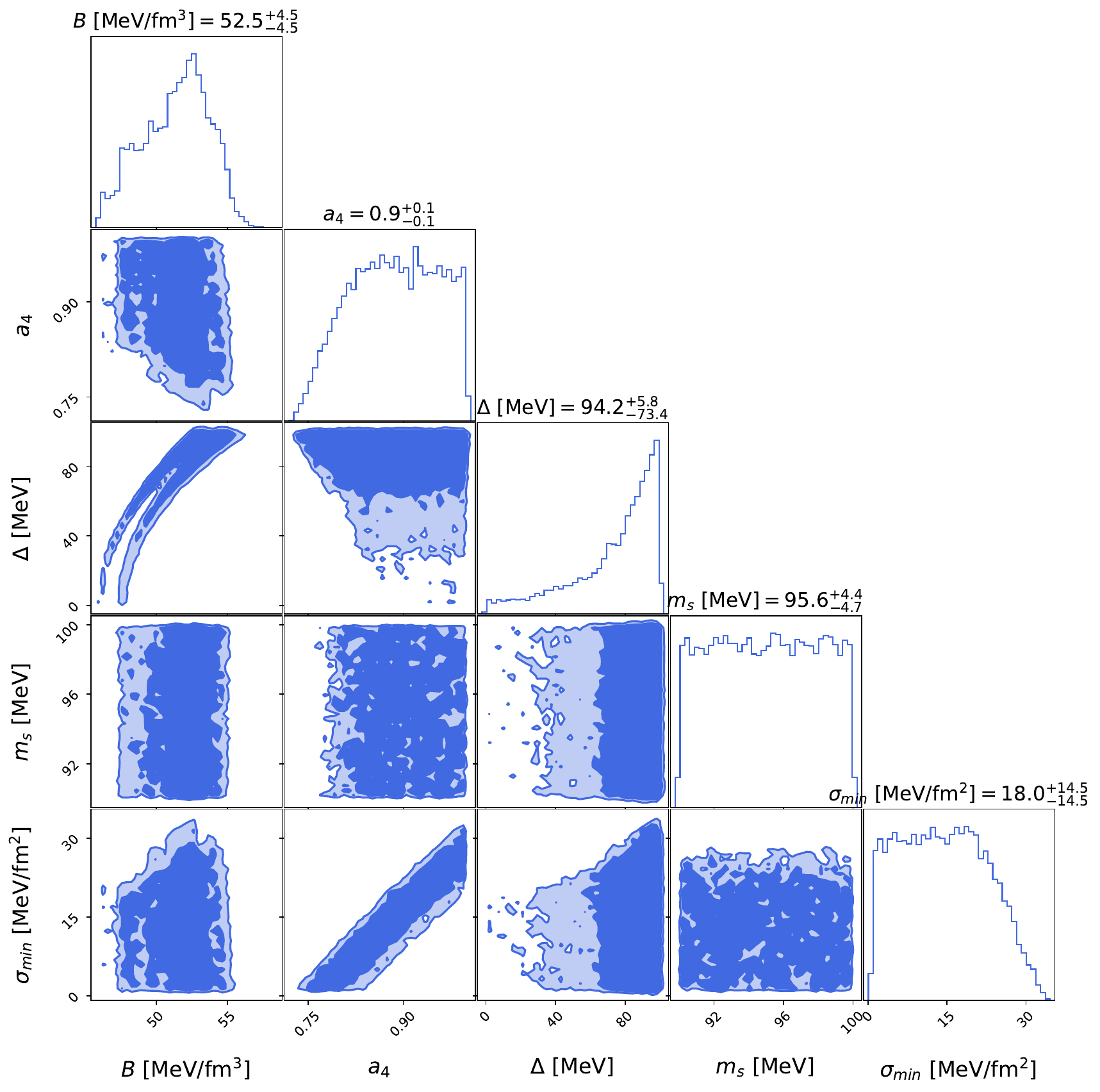}
  \caption{(Color online) Posterior probability distribution functions of the five parameters for the QM properties and their correlations at 68\% (light shadow) and 90\% (dark shadow) confidence levels inferred from the Bayesian analysis of the data listed in Table \ref{tab-data}. In the calculations, the Prior-1 set is used. The values showed in the corner are the 90\% confidence intervals for the five parameters.}\label{fig:5paraCor-pr1}
\end{center}
\end{figure*}
\begin{figure*}[ht]
\begin{center}
  \includegraphics[width=15cm,height=12cm]{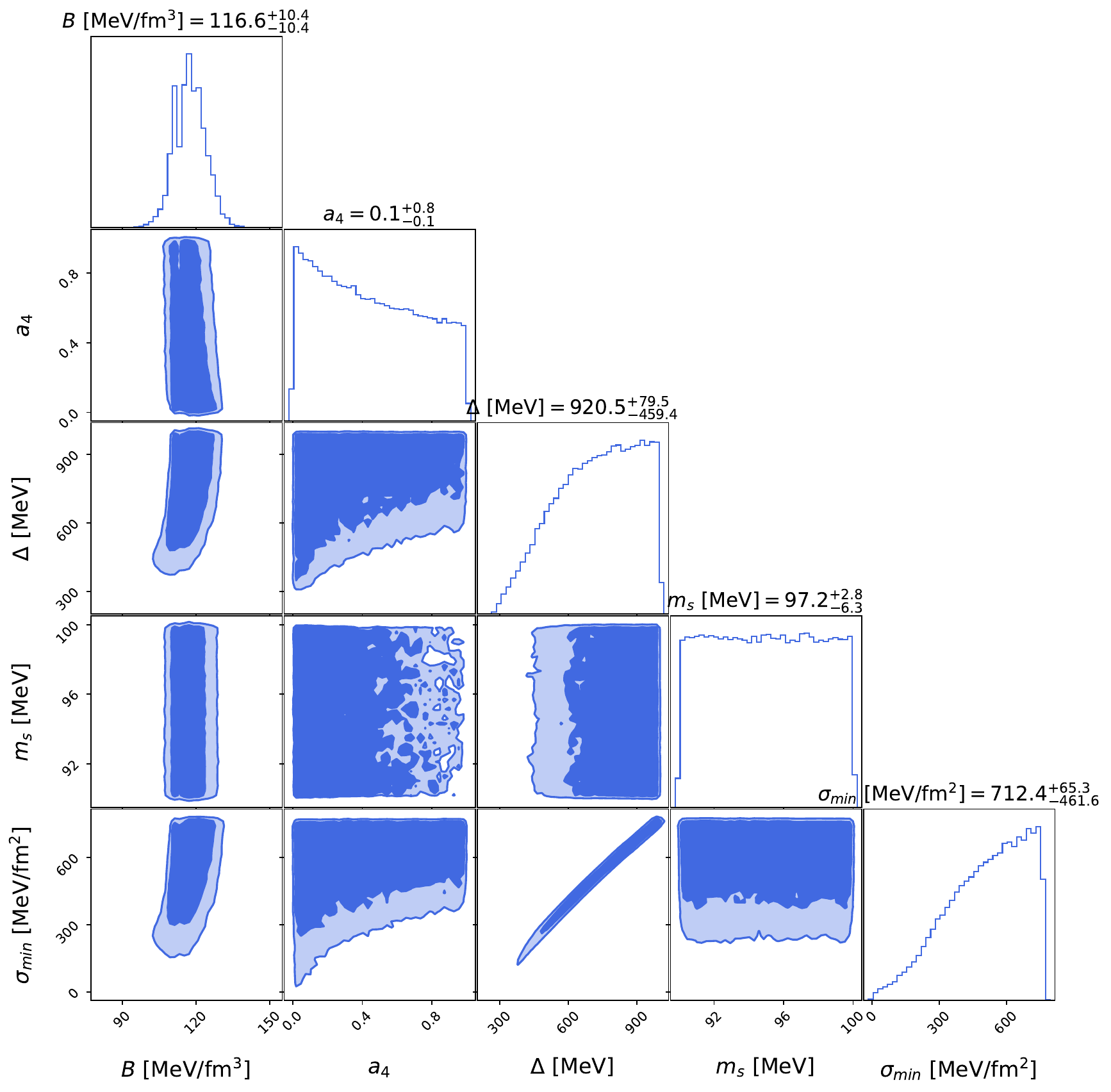}
  \caption{(Color online) Same as Fig. \ref{fig:5paraCor-pr1}, but showing the results corresponding to the Prior-2 prescription.}\label{fig:5paraCor-pr2}
\end{center}
\end{figure*}
Figures~\ref{fig:5paraCor-pr1} and ~\ref{fig:5paraCor-pr2} present the posterior distributions of the four parameters, along with the derived minimum surface tension, obtained using the four-parameter quark star model and inferred via Bayesian statistical methods for the 2SC quark phase. The Bayesian analysis incorporates the observational data summarized in Table \ref{tab-data}. The lighter and darker shaded regions correspond to the 90\% and 68\% credible intervals, respectively. The diagonal elements of the figure display the 90\% credible intervals for individual parameters.

As illustrated, and consistent with previous studies~\cite{Miao2021_ApJ917-L22,Wang2024_JCAP2024-038}, the current observational data impose stringent constraints on the bag parameter $B$. Under the Prior-1 configuration, the 90\% credible interval for $B$ is determined to be 52.5 $\pm$ 4.5 $\text{ MeV/fm}^3$, which is consistent with the result of 46.08$^{+6.8}_{-5.65}$ $\text{ MeV/fm}^3$ (90\% credible interval) reported in Ref.~\cite{Miao2021_ApJ917-L22} and $48.83^{+2.43}_{-1.94}$ $\text{ MeV/fm}^3$ (68\% credible interval) reported in Ref.~\cite{Wang2024_JCAP2024-038}. When the Prior-2 configuration is adopted, the inferred value becomes 116.6 $\pm 10.4 \text{ MeV/fm}^3$ at 90\% confidence level, a result that supports the proposition in Ref. ~\cite{Miao2021_ApJ917-L22} that interpreting the GW190814 secondary component as a quark star would require $B$ in the range of $B^{1/4}$= 170-192 MeV ($B$=109–178 $\text{ MeV/fm}^3$), and our analysis yields a more tightly constrained estimate.

With the Prior-1 setup, a relatively large value of $a_4 = 0.9 \pm 0.1$ (90\% credible interval) is found, indicating that the current observational data favor smaller QCD corrections. In contrast, the Prior-2 setup yields an opposite trend, suggesting that the constraint on $a_4$ is sensitive to the prior ranges adopted for parameters $B$ and $\Delta$. For the pairing gap parameter $\Delta$, the Prior-1 configuration uses a range of 0–100 MeV, consistent with that in Ref. ~\cite{Miao2021_ApJ917-L22}, while Prior-2 extends this to 0–1000 MeV, aligning with the range used in Ref. \cite{Kurkela2024_PRL132-262701}. Our results indicate that current data provide only a weak constraint on this parameter. A similarly weak constraint is found for the strange quark mass $m_s$. Depending on the prior configuration, the minimum surface tension $\sigma_{\text{min}}$ is inferred to be either relatively small, 18 $\pm 14.5 \text{ MeV/fm}^2$ (Prior-1), or significantly larger, 712$^{+65.3}_{-461.6} \text{ MeV/fm}^2$ (Prior-2), both at the 90\% credible level.

Under the Prior-1 configuration, pronounced positive correlations are observed between parameter $B$ and $\Delta$ as well as between $a_4$ and $\sigma_{\text{min}}$ for the 2SC phase, which are attributed to both the stability and astrophysical constraints. However, these correlations diminish when the Prior-2 setup is used. Furthermore, a strong positive correlation emerges between $\sigma_{\text{min}}$ and $\Delta$ when Prior-2 is employed, a feature not present in the Prior-1 results and is attributed to both the stability constraint indicated in Eq.~(\ref{eq:epa_266Hs}). Correlations among the remaining parameters are generally weak for both prior assumptions.

\begin{figure*}[ht]
\begin{center}
  \includegraphics[width=15cm,height=10cm]{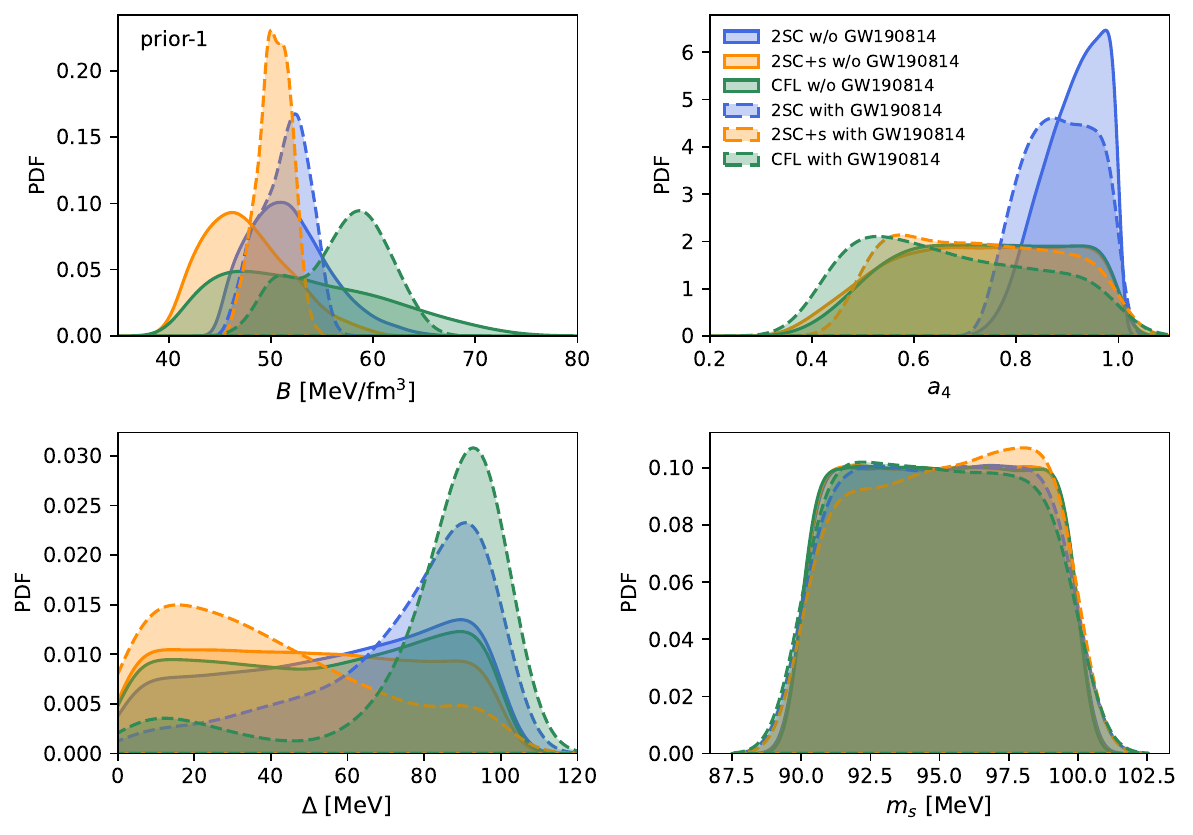}
  \caption{(Color online) Posterior probability distribution functions of the radius, tidal deformability corresponding to the quark stars with 1.4 and 2.0 solar masses, maximum mass and the corresponding radius, central squared speed of sound, central energy density and central pressure inferred from the Bayesian analysis of the data listed in Table \ref{tab-data} for the CFL, 2SC and 2SC+s phases.}\label{fig:4paraPDF-pr1}
\end{center}
\end{figure*}

\begin{figure*}[ht]
\begin{center}
  \includegraphics[width=15cm,height=10cm]{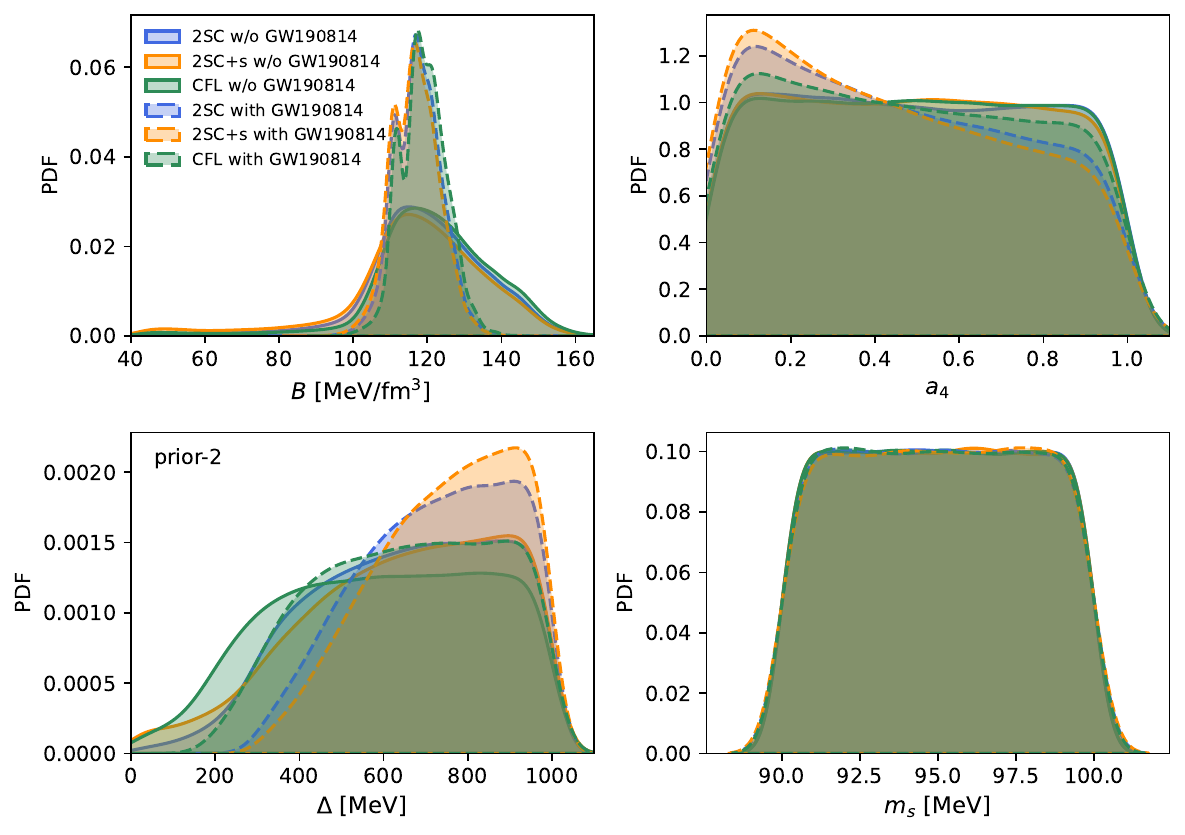}
  \caption{(Color online) Posterior probability distribution functions of the radius, tidal deformability corresponding to the quark stars with 1.4 and 2.0 solar masses, maximum mass and the corresponding radius, central squared speed of sound, central energy density and central pressure inferred from the Bayesian analysis of the data listed in Table \ref{tab-data} for the CFL, 2SC and 2SC+s phases.}\label{fig:4paraPDF-pr2}
\end{center}
\end{figure*}

Figures~\ref{fig:5paraCor-pr1} and ~\ref{fig:5paraCor-pr2} present the Bayesian inference results for the four parameters characterizing the 2SC quark phase under the two prior choices, Prior-1 and Prior-2. The posterior distributions of the parameters describing the other two strange quark phases, namely 2SC+s and CFL, are shown in Figures \ref{fig:4paraPDF-pr1} and \ref{fig:4paraPDF-pr2}, respectively. For each prior setup, we not only compare the constraints imposed by observational data on the parameters of the three quark phases, but also investigate the impact of the secondary component of GW190814 on the inferred posteriors.

Under the Prior-1 setup (Fig. \ref{fig:4paraPDF-pr1}), when the GW190814 secondary constraint is not included (solid curves), the bag parameter associated with the 2SC phase tends to take relatively larger values, whereas these of the CFL phase is much more weakly constrained. This behavior indicates a softer EOS for quark matter in the 2SC phase, while the CFL phase allows for a broader range of stiffness. Owing to the stability constraint indicated in Eq.~(\ref{eq:epa_266Hs}), the observational data provide a strong constraint on the parameter ($a_4$) for the 2SC phase, while being largely insensitive to these for the two strange quark phases. In contrast, the posterior distributions of $\Delta$ and $m_s$ are nearly identical for all three quark phases.

When the GW190814 secondary constraint is taken into account (dashed curves), the bag parameters of all quark phases become significantly better constrained. Among them, the CFL phase favors the largest bag parameter, followed by the 2SC phase, while the 2SC+s phase corresponds to the smallest one. The posterior distributions of $a_4$ remain essentially unchanged compared to the case without GW190814. For the pairing gap ($\Delta$), the most probable values for the 2SC and CFL phases are relatively large, whereas the 2SC+s phase prefers a smaller value. This feature arises from the fact that, under the Prior-1 setup, the parameter $\lambda$ is allowed to take negative values, i.e., Eq. (\ref{eq:lam}) can become negative, and the GW190814 observation provides a non-negligible constraint on $\Delta$. By contrast, the strange quark mass ($m_s$) shows only minor differences between the cases with and without GW190814.

Under the Prior-2 setup (Fig. \ref{fig:4paraPDF-pr2}), some qualitative features remain consistent with those obtained using Prior-1: GW190814 still imposes a strong constraint on the bag parameter, while having little impact on $m_s$. However, notable differences also emerge. In particular, the bag parameters of the three quark phases are almost identical regardless of whether the GW190814 constraint is included, and the effect of surface tension on the parameter $a_4$ becomes negligible. Although GW190814 provides a certain level of constraint on the pairing gap ($\Delta$), the resulting posterior distributions show only minor differences among the three quark phases.
\begin{figure*}[ht]
\begin{center}
  \includegraphics[width=15cm,height=12cm]{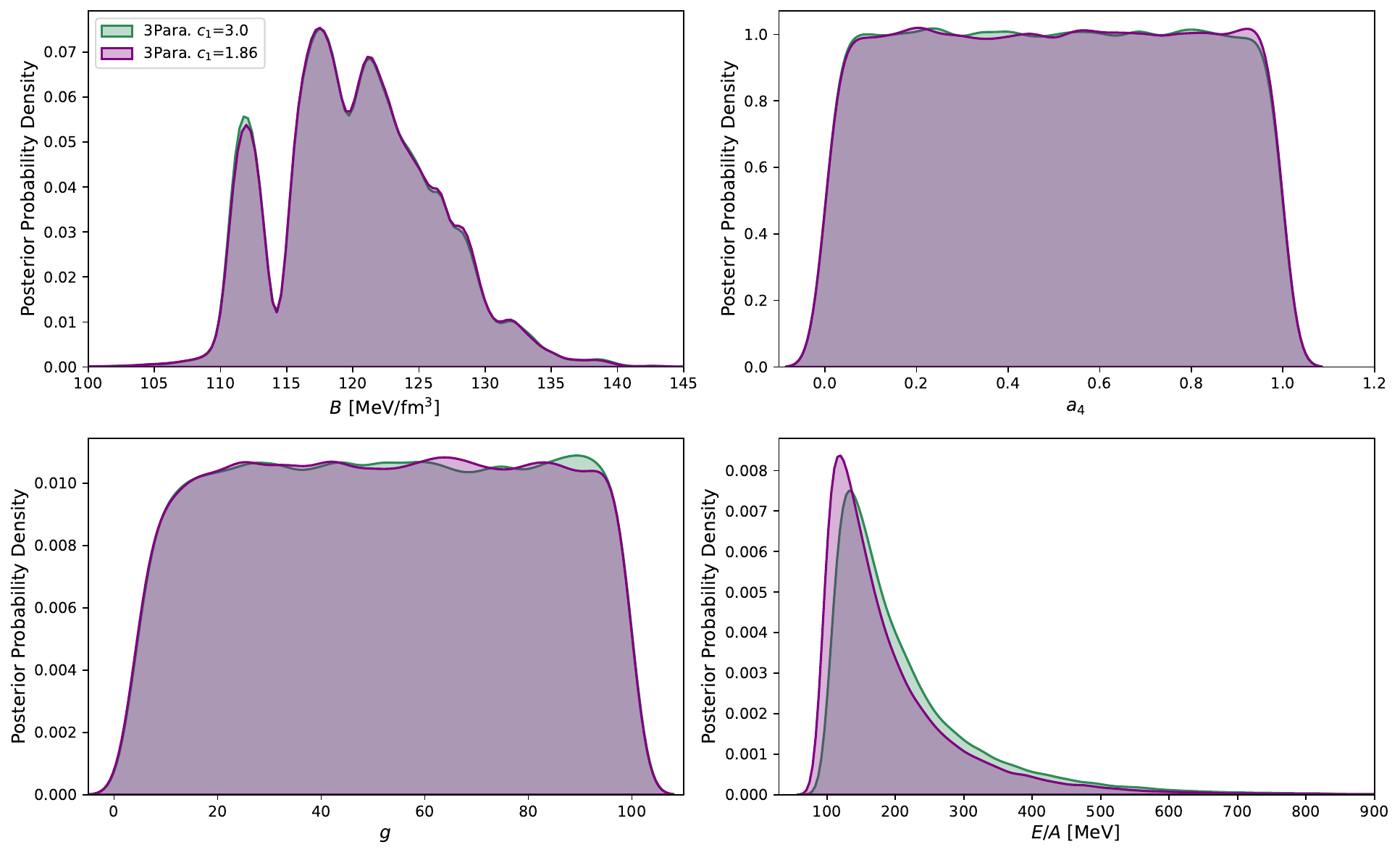}
  \caption{(Color online) Posterior probability distribution functions of the three parameters for the three-parameter QS model and the energy per baryon number inferred from the Bayesian analysis of the data listed in Table \ref{tab-data}. }\label{fig:3para-pr2}
\end{center}
\end{figure*}
\begin{figure}[ht]
\begin{center}
  \includegraphics[width=8cm,height=8cm]{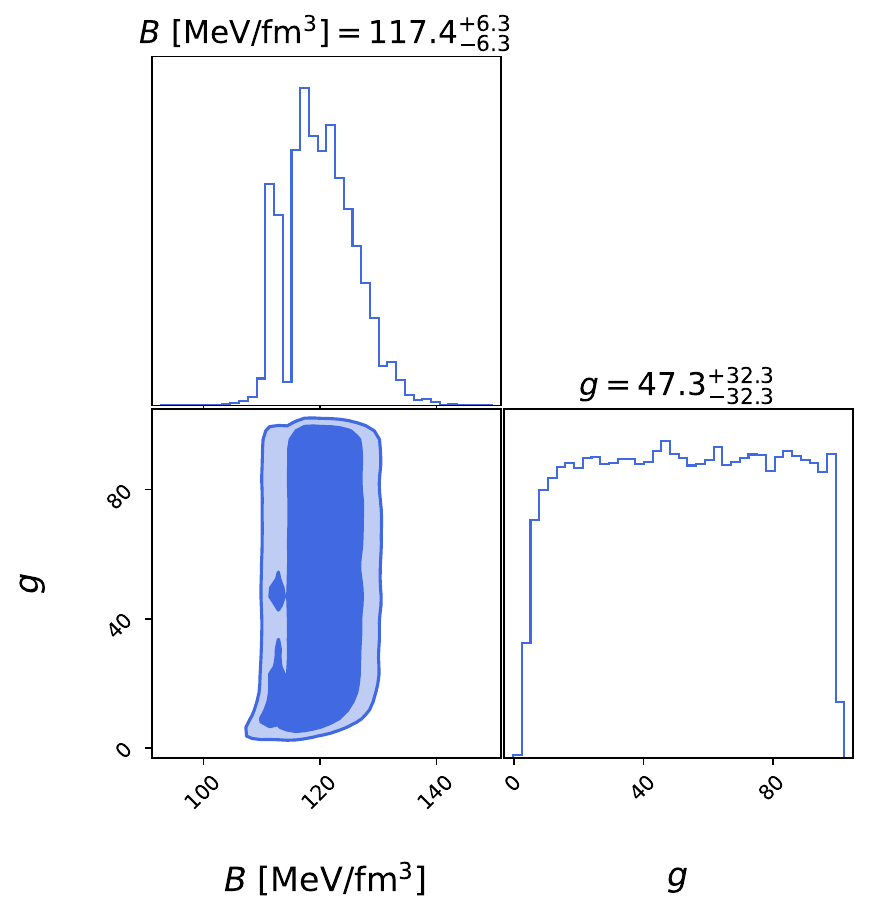}
  \caption{(Color online) Posterior probability distribution functions of the two parameters for the two-parameter QS model and their correlations at 68\% (light shadow) and 90\% (dark shadow) confidence levels inferred from the Bayesian analysis of the data listed in Table \ref{tab-data}. The values showed in the corner are the 90\% confidence intervals for the two parameters.}\label{fig:2paraCor}
\end{center}
\end{figure}

 \begin{figure*}[ht]
\begin{center}
  \includegraphics[width=15cm,height=10cm]{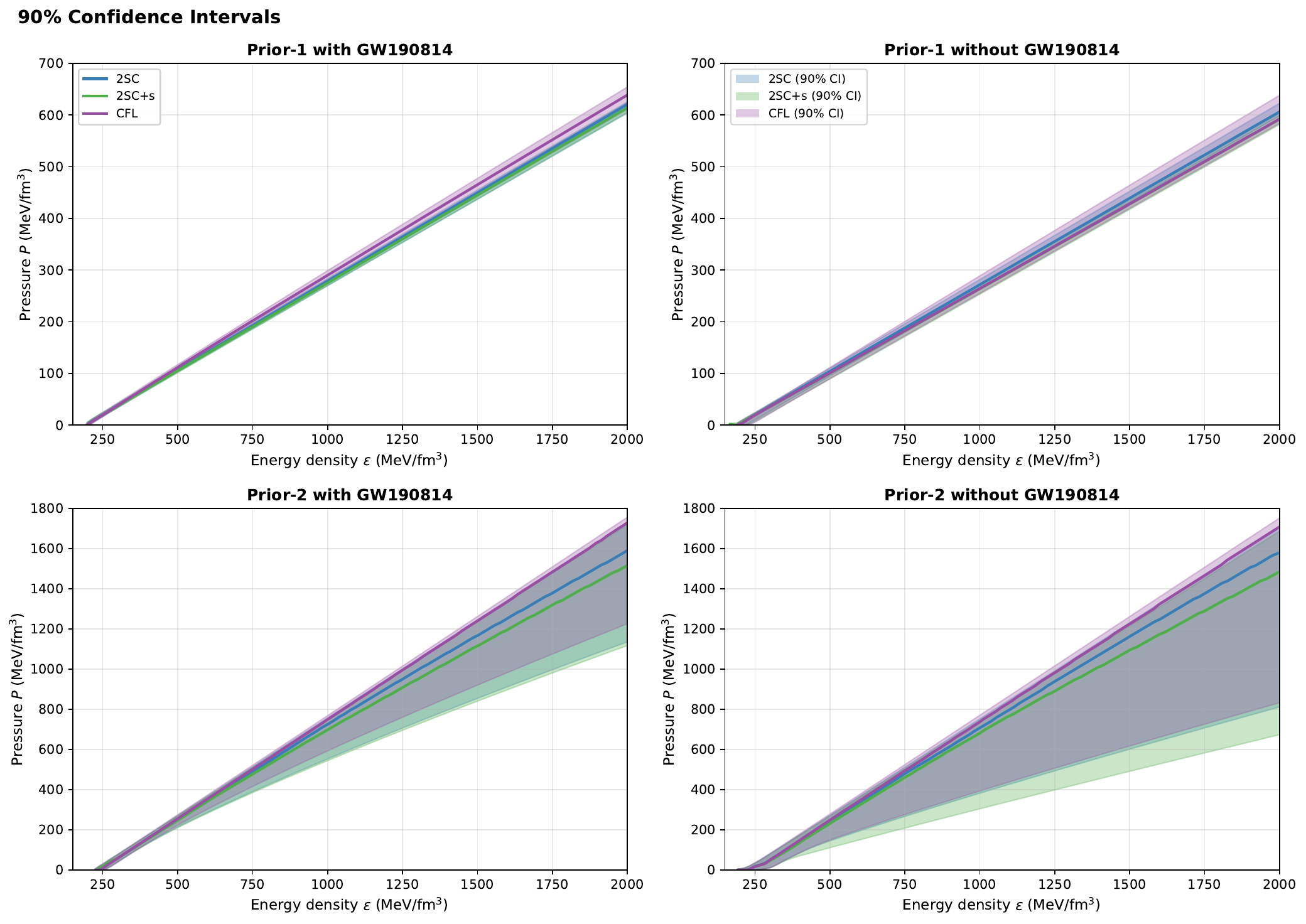}
  \caption{(Color online) Posterior probability distribution functions of the radius, tidal deformability corresponding to the quark stars with 1.4 and 2.0 solar masses, maximum mass and the corresponding radius, central squared speed of sound, central energy density and central pressure inferred from the Bayesian analysis of the data listed in Table \ref{tab-data} for the CFL, 2SC and 2SC+s phases.}\label{fig:pres-eden}
\end{center}
\end{figure*}
\begin{figure*}[ht]
\begin{center}
  \includegraphics[width=15cm,height=10cm]{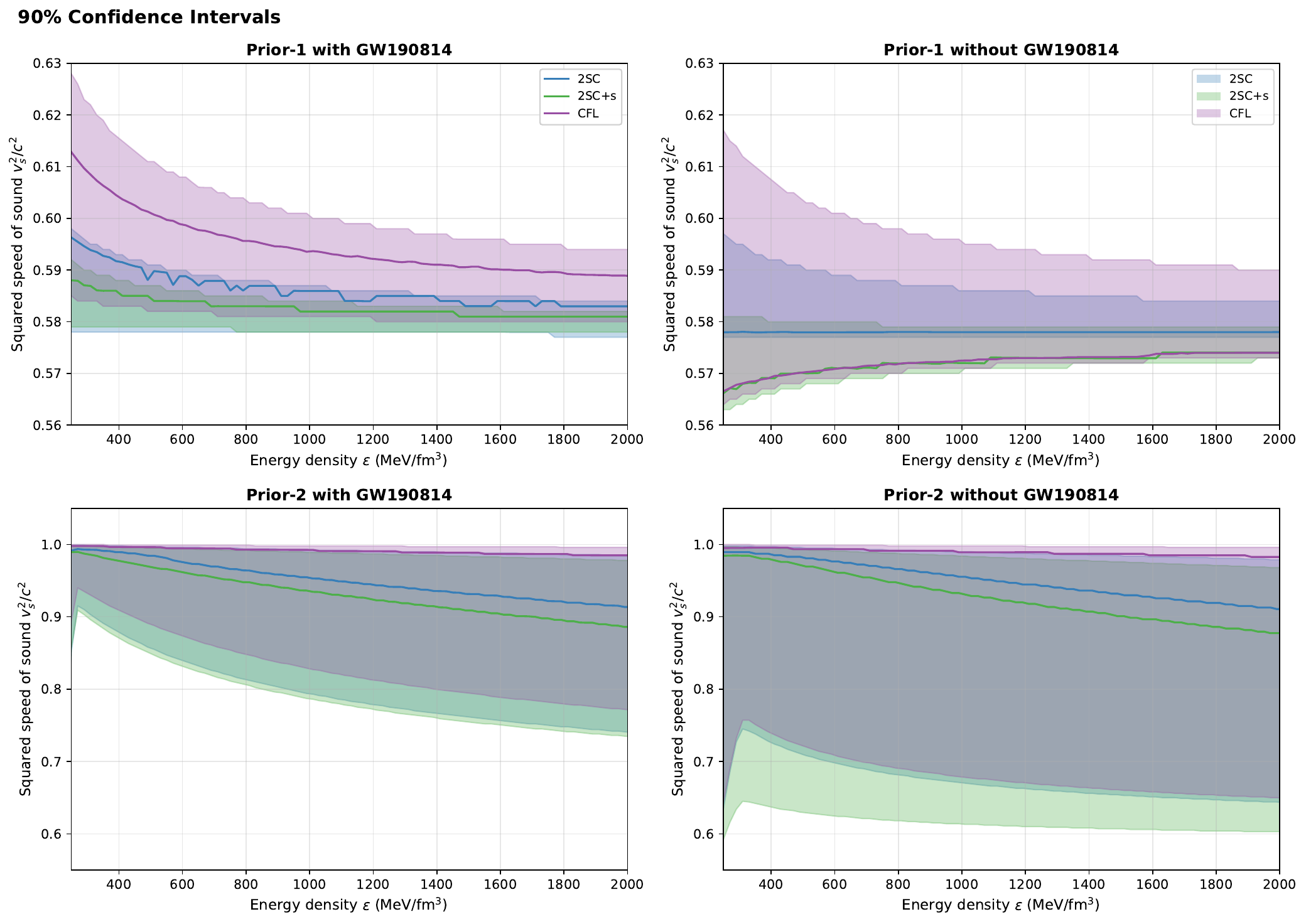}
  \caption{(Color online) Posterior probability distribution functions of the radius, tidal deformability corresponding to the quark stars with 1.4 and 2.0 solar masses, maximum mass and the corresponding radius, central squared speed of sound, central energy density and central pressure inferred from the Bayesian analysis of the data listed in Table \ref{tab-data} for the CFL, 2SC and 2SC+s phases.}\label{fig:vs-sound}
\end{center}
\end{figure*}

\subsection{\label{sec:2paraPDF}Reduced quark star models: three- and two-parameter descriptions}
To assess the necessity of the full four-parameter description, we next examine reduced quark star models.
Figure \ref{fig:3para-pr2} presents the Bayesian inference results obtained within the framework of the three-parameter quark star model, showing the posterior distributions of the parameters $B$, $a_4$, and  $g$, as well as the corresponding transformed quantity, the energy per baryon number ($E/A$) of quark matter, for different choices of $c_1$. Here, $c_1 = 1.86$ corresponds to the 2SC quark phase, while $c_1 = 3$ is adopted for the two strange quark phases, 2SC+s and CFL\cite{Zhang2021_PRD103-063018}. The prior ranges of the three parameters are chosen as 0-200~$\mathrm{MeV/fm^3}$ for $B$, 0-1 for $a_4$, and -100-100 for $g$.

As shown in Fig. \ref{fig:3para-pr2}, the posterior distributions of the three parameters, as well as that of $E/A$, remain almost unchanged for different values of $c_1$. This indicates that the impact of the parameter $c_1$ can be safely neglected when employing the three-parameter quark star model. Moreover, the Bayesian inference results for $B$ and $a_4$ are in good agreement with those obtained from the four-parameter quark star model. Although the current observational data provide only weak constraints on $g$, its posterior range is still reduced to 0-100 compared to the adopted prior interval (-100-100).
Within the adopted three-parameter prior ranges, the posterior distribution of the energy per baryon is found to lie approximately in the range 50-700 MeV. Consequently, the condition ($E/A < 930~\mathrm{MeV}$), i.e., the stability criterion given by Eq. (\ref{eq:epa_max}), is always satisfied. This implies that the stability condition does not need to be explicitly imposed when using the three-parameter quark star model.

In light of these results, the three-parameter quark star model can be further reduced to a two-parameter model, characterized solely by $B$ and $g$. In the Bayesian inference based on this two-parameter model, the prior ranges of both parameters are kept identical to those in the three-parameter case. The resulting posterior distributions and their correlations are shown in Fig. \ref{fig:2paraCor}. We find that the posteriors of $B$ and $g$ are consistent with those obtained in the three-parameter model, while the constraint on $B$ becomes significantly tighter compared to the four-parameter model, as indicated in Figs. \ref{fig:5paraCor-pr2} and \ref{fig:3para-pr2}, respectively. For instance, the four-parameter model yields $B = 116.6 \pm 10.4~\mathrm{MeV/fm^3}$, whereas the two-parameter model gives $B = 117.4 \pm 6.3~\mathrm{MeV/fm^3}$.
It should be emphasized that the two-parameter model adopted in the present work is fundamentally different from the two-dimensional (2-d) model introduced in Ref. \cite{Wang2024_JCAP2024-038}. The latter is applicable only to the CFL phase, whereas the two-parameter model proposed here is valid not only for the CFL phase but also for the 2SC and 2SC+s phases, thereby exhibiting a broader range of applicability.

\subsection{\label{sec:eos-vs}Equation of state and sound speed}
Figures \ref{fig:pres-eden} and \ref{fig:vs-sound}, respectively, display the 90\% credible regions of the pressure ($P$) and squared sound speed ($v_s^2/c^2$) as functions of energy density for quark stars based on the four-parameter quark matter model.
In both figures, the upper and lower panels correspond to the results obtained with the Prior-1 and Prior-2 setups, respectively, while the left and right panels illustrate the cases with and without the inclusion of the GW190814 secondary mass constraint in the Bayesian inference. The solid curves in the figures denote the most probable values for the pressure and squared speed of sound at the given energy densities.

As can be seen from the figures, the quark matter EOS occupies a relatively narrow band, reflecting the strong constraints imposed by the observational data. The broader Prior-2 allows a much wider range of EOS behavior, particularly at high densities, consistent with the extended maximum-mass range inferred in this case.
In all scenarios, the EOSs corresponding to different quark phases are nearly identical, reinforcing the conclusion that current data are insensitive to the detailed phase structure of quark matter.
It is worth emphasizing that the mass measurement of the GW190814 secondary component imposes strong constraints on the quark star EOS, significantly restricts the allowed parameter space of the quark star EOS.

There have been numerous prior discussions suggesting that, in order to satisfy the lower-mass limit of two solar masses for compact stars, the value of $v_s$ in compact-star matter should significantly surpass the conformal limit of c/$\sqrt{3}$~\cite{Kurkela2014_ApJ789-127, Bedaque2015_PRL114-031103, Alsing2018_MNRAS478-1377, Cai2023_PRD108-103041}. In some studies, i.e., see~\cite{Tews2018_PRC98-045804}, it has even been observed to be close to 0.9$c$ around 5$n_0$. These results align with theoretical predictions from perturbative QCD~\cite{Xia2021_CPC45-055104} and Bayesian analyses incorporating multi-messenger data~\cite{Li2021_ApJ913-27, Landry2020_PRD101-123007}.
In the present work, our calculations also show that the sound speed exceeds the conformal limit, as illustrated in Fig. \ref{fig:vs-sound}.
It is found that the behavior of the sound speed closely mirrors the corresponding EOS characteristics discussed in Fig. \ref{fig:pres-eden}. Under the Prior-1 setup, the quark matter sound speed exhibits a relatively narrow band, reflecting the limited variation of the quark EOS,
and the Prior-2 setup permits a much broader range of sound speeds for quark matter, in agreement with the wider EOS uncertainty inferred under this prior choice. In all cases, the three quark phases show very similar sound-speed behavior, reinforcing the conclusion that their EOSs are nearly indistinguishable within the present Bayesian framework. Consistent with Fig. \ref{fig:pres-eden}, the GW190814 constraint also strongly bounds the squared sound speed, significantly narrowing its allowed range, an effect that is clearly evident when using the Prior-2 configuration.

\section{\label{sec:con}Conclusion and Outlook}
In this work, we have performed a comprehensive Bayesian analysis of the quark star
equations of state using current multimessenger observations, with particular emphasis
on assessing the impact of prior assumptions and the role of extreme-mass constraints.
By systematically comparing two prior prescriptions and analyses with and without
including the GW190814 secondary component, we have quantified both robust and
prior-dependent aspects of the inferred properties of quark stars.

We find that several macroscopic properties of quark stars are tightly constrained by
existing observations and remain largely insensitive to prior choices.
In particular, the inferred radii at canonical masses, the overall compactness, and the
compatibility with low-mass compact objects such as HESS~J1731--347 emerge as robust
features across all considered quark phases and model parameterizations.
These results highlight the distinctive role of self-bound quark matter in naturally
accommodating compact low-mass configurations.

In contrast, properties associated with the extreme-density regime, including the
maximum mass, the high-density stiffness of the equation of state, and the behavior of
the sound speed, exhibit pronounced prior dependence.
The inclusion of the GW190814 secondary component significantly reshapes the inferred
posteriors by favoring stiffer quark matter equations of state, but this effect is
strongly contingent on the assumed prior ranges.
Consequently, GW190814 should be interpreted as a powerful diagnostic of the allowed
high-density parameter space rather than as a definitive confirmation of the existence
of quark stars.

At the microscopic level, current multimessenger data primarily constrain the bulk
properties of quark matter, most notably the effective bag parameter, while leaving
other parameters related to quark pairing, strange quark mass, and surface tension
largely unconstrained.
This indicates that present observations are sensitive mainly to the macroscopic
stiffness of quark matter, with limited discriminatory power regarding its detailed
phase structure.

Our results underscore the importance of transparent prior selection and explicit
prior-sensitivity analyses in Bayesian studies of dense matter.
Future observations offer several promising avenues to further test the quark star
hypothesis.
High-precision radius measurements of low-mass compact stars would provide a
particularly decisive probe of self-bound matter, while improved gravitational-wave
constraints on tidal deformabilities and the discovery of additional extreme-mass
compact objects could substantially reduce the prior dependence of inferences in the
high-density regime.
Together, such observations will be essential for establishing whether quark stars are
realized in nature or whether current consistency reflects the remaining flexibility of
dense-matter models.

It is also interesting and important to perform a Bayesian factor comparison of quark stars and neutron stars~\cite{Cao:2023rgh} incorporating the more updated observation data we explored here. Besides, quark stars may develop a hybrid core of different quark phases, forming so-called hybrid quark stars~\cite{Yang:2025iyv,Zhang:2025rnf}. Bayesian analyses on these new objects are also worth further explorations. We leave these for future studies.

\section*{ACKNOWLEDGMENTS}
C.J.X. is supported by the National Natural Science Foundation of China (Grant No. 12275234) and the National SKA Program of China (Grant No. 2020SKA0120300). W.J.X. is supported by the Open Project of Guangxi Key Laboratory of Nuclear Physics and Nuclear Technology under Grant No. NLK2023-03, and the Central Government Guidance Funds for Local Scientific and Technological Development, China, under Grant No. Guike ZY22096024. C.Z. is supported by the Fundamental Research Funds for the
Central Universities. R.X.X. is supported by the National SKA Program of China  (Grant No. 2020SKA0120100).

\newpage

%

\end{document}